\documentclass[oldversion]{aa}

\usepackage{graphicx}
\usepackage{natbib}
\bibpunct{(}{)}{;}{a}{}{,}
\usepackage{amssymb}

\newcommand{\grad}{\vec{\nabla}}
\newcommand{\dive}{\vec{\nabla}.}

\begin{document}

\title{A Viscous Heating Mechanism for the Hot Plasma in the Galactic Center Region}
\titlerunning{Viscous heating in the GC}
\author{R. Belmont\thanks{\email{belmont@cea.fr}} \and M. Tagger}
\institute{CEA Service d'Astrophysique, UMR "AstroParticlues et Cosmologie", Orme de merisiers, 91191 Gif-sur-Yvette, France }
\date{Received --- / Accepted ---}

\abstract{
In addition to lines originating in a soft phase at $\sim$ 0.8 keV and to cold molecular clouds, the X-ray spectra from the Galactic center region also exhibit properties similar to those of a diffuse, thin, very hot plasma at 8 keV on a scale of hundreds of parsecs. This phase is surprising for more than one reason. First, such a hot plasma should not be bound to the Galactic plane and the power needed to sustain the escaping matter would be higher then any known source. Second, there is no known mechanism able to heat the plasma to more than a few keV. Recently we have suggested that, hydrogen having escaped, the hot plasma could be a helium plasma, heavy enough to be gravitationally confined. In this case, the required power is much more reasonable.
We present here a possible heating mechanism which taps the gravitational energy of the molecular clouds. We note that the 8 keV plasma is highly viscous and we show how viscous friction of molecular clouds flowing within the hot phase can dissipate energy in the gas and heat it. 
We detail the MHD wake of a spherical cloud by considering the different MHD waves the cloud can excite.
We find that most of the energy is dissipated by the damping of Alfv\'enic perturbations in two possible manners, namely by non-linear effects and by a large scale curvature of the field lines. We find that the total dissipation rate depends on the field strength. For fields B~$\lesssim 200\mu$G both mechanisms produce power comparable to or higher than the radiative losses; for strong fields B~$\gtrsim$~1 mG, only the curvature damping can balance the X-ray emission and requires a radius of curvature $R_\mathrm{c}\lesssim$ 100 pc; whereas for intermediate fields, the total dissipation is more than one order of magnitude smaller, requiring a higher accretion rate. We note that the plasma parameters may be optimal to make the dissipation most efficient, suggesting a self-regulation mechanism. The loss of kinetic and gravitational energy also causes accretion of the clouds and may have significant action on the gas dynamics in this region between the large scale, bar dominated flow and the central accretion to the massive black hole.
 }
\maketitle
\section{Introduction}
The optical emission from the Galactic center is strongly obscured by dust absorption. Radio, IR and X-ray astronomy have thus become useful tools to probe this region.  The present picture of the center region is very complex, showing supernova remnants, H$_2$ regions, star clusters, extended filaments, magnetism, star forming regions... For more than 20 years, many high energy observations have reported an intense X-ray emission from the Galactic center region \citep{Worall82, Warwick85}. Although a cosmic contribution has been identified as resulting from discrete sources, the remaining Galactic ridge emission has not been resolved yet. Some spectral properties are very similar to those of Galactic X-ray point sources \citep{Wang02} but recent Chandra observations have shown that no more than $10\% $  of the emission can be attributed to point sources \citep{Ebisawa01, Muno04} \citep[see however][for a different vue]{Revnivtsev05}. 

The Galactic ridge X-ray emission may thus be truly diffuse and its spectrum is a powerful diagnostic to understand its nature. In addition to the bremstrahlung emission, an intense highly ionized He-like Fe emission line at 6.7 keV was first reported by Ginga \citep{Koyama86a}. The continuum and line are observed to extend out to several kpc \citep{Yamauchi96} with a typical scale height of less than 100 pc, but a strong enhancement is observed in the inner 300 pc \citep{Yamauchi90}. These observations have then been confirmed by many others and new H- and He-like lines from very ionized elements have been resolved with ASCA in the $.5-10$~keV band \citep{Kaneda97}. The study of these Mg, Si, S, Ar, Ca and Fe lines indicates that the spectrum cannot be modeled with a single-temperature thermal plasma. As a result two classes of models have been proposed. 

On the one hand, it has been proposed that the spectrum could result from a single-temperature plasma associated with non- or quasi-thermal mechanisms \citep{Valinia00, Tanaka00, Masai02, Dogiel02b, Dogiel02a}. However, besides the intrinsic problems of the different models, recent observations with the Chandra satellite have  provided spectra more consistent with a thermal origin \citep[see][for a review of these models and a comparison with results of Chandra observations]{Muno04}. 

On the other hand, authors have interpreted the spectrum as originating from two different spectral components \citep{Kaneda97, Muno04}: a soft component at kT $\sim$ 0.8 keV and a hot component at kT $\sim$ 8 keV. The soft plasma properties are compatible with a supernova origin: its spatial distribution is patchy and supernova shock-waves are known to generate temperatures of $\sim$ 1 keV. 

The origin of the more diffuse 8 keV plasma is puzzling and raises several questions. Assuming that the plasma has solar abundances, i.e. is essentially made of hydrogen, it has been noted that its temperature is too high for it to be gravitationally bound to the Galactic plane \citep{Koyama96, Muno04}. The power required to compete with the energy losses associated with this escaping matter exceeds the power of  any known source. An average supernova rate $10^2$ or $10^4$ higher than in the rest of the Galaxy would for instance be required to heat the hot plasma before it escapes the central region \citep[derived from][]{Muno04}. 

More recently, we have noticed that, at the inferred temperature and density of the hot phase, hydrogen ions are weakly collisional with other ions so that they can leave the Galactic plane without dragging other elements with them, leaving a helium plasma that is heavier and confined by gravity \citep{Belmont05}. As this plasma does not escape, energy losses are dominated by radiation which occurs on much longer time scales: $\sim 10^8$~yr \citep{Muno04}. 
From the Ginga data of \citet{Yamauchi90} we derive a peak emissivity of 1.4-4.6$\times10^{33}$ erg s$^{-1}$arcmin$^{-2}$ in the central region. This result is rather consistent with recent observations by Chandra of the inner 20 pc which have reported a local  luminosity of L=5.-9.$\times 10^{33}$ erg s$^{-1}$arcmin$^2$ \citep{Muno04}. The Ginga value is an average over a large area whereas the Chandra field is more central and much smaller. This could explain the slight difference. With the intermediate value, the total emission in the whole central region is L $\approx4.\times10^{37}$ erg s$^{-1}$.
This power can be provided by reasonable sources. Nevertheless, a mechanism must still be found that can heat the gas up to 8 keV. Supernovae for example could provide enough energy to balance the radiative cooling, but, although their temperature depends on their age and the external pressure, SNRs have not been observed to temperature higher than $\sim$ 1-3 keV after a few hundred years, significantly smaller than the required 8 keV. 

In this paper, we assume that the 8~keV emission indeed originates in a diffuse helium plasma in the Galactic center and we investigate the idea that the heating in this region could be provided by the friction of cold clouds with the surrounding gas. This idea is based on two main facts. 

First, many molecular clouds have been observed  in CO, CS, NH$_3$ in the central region. The dense H$_2$ clouds of about 10~pc size are observed to form in a ring at about 180~pc from Sgr~A*, then detach and spiral inward with a significant velocity relative to the surrounding medium, typically 100~km~s$^{-1}$.  The large number of moving clouds represents a huge reservoir of gravitational and kinetic energy that may be used to heat the plasma.

Second, plasmas at temperatures as high as 8 keV can be highly viscous \citep{Braginskii65}. This viscosity is however different from the usual one characteristic of neutral gases. Indeed, in the conditions of the Galactic center, any reasonable magnetic field fully inhibits the usual shear viscosity and so reduces by orders of magnitude the efficiency of the corresponding viscous dissipation. However, we will show that the remaining bulk viscosity, which acts on compressional motions, is sufficient to apply a significant net viscous stress on the moving clouds. The associated power is dissipated in the plasma and heats it. 

The present paper is organized as follows:  first, we describe in sect. \ref{arena} the Galactic center arena, its molecular and magnetic content; then, we present the main characteristics of the viscosity acting in the Galactic center in sect. \ref{viscosity}; we describe how viscous friction with cold molecular clouds can dissipate part of the cloud's kinetic energy in sect. \ref{heating}, and last, we discuss the efficiency of this heating in providing the energy required to heat the medium in sect. \ref{discuss}.

\section{The Central Molecular Zone}
\label{arena}
The Central Molecular Zone \citep[CMZ,][]{Morris96} covers the first hundreds of parsecs and differs from the outer part of the Galaxy by many features. Diffuse radio, IR and X-ray emission are for example higher there than anywhere else \citep{Brogan03,Bennett94,Yamauchi90}. The molecular content of this region is higher than in the outer parts of the Galaxy and about 10\% of the Galactic molecular gas in condensed in the CMZ. The outer limit is composed by a dense torus of gas at about 180-200~pc. At this radius, the gas density drops from 200 M$_\odot$pc$^{-2}$ to 5 M$_\odot$pc$^{-2}$. Last but not least, many non-thermal filaments are observed in the first 300~pc, which are not observed anywhere else in the Galaxy.  In this paper, we focus on a region of 150~pc radius and $70$~pc high below and above the disk, which represents the main volume of the Central Molecular Zone \citep{Morris96} but does not include the dense torus. We will refer to the 'inner CMZ' for this region, which corresponds roughly to the region where the filaments are observed. We assume that the distance to Sgr~A* is 8.5~kpc.

Our model relies on tapping some of the kinetic energy of the numerous molecular clouds in the CMZ. The statistical properties (size, mass, internal velocity dispersion...) of these H$_2$ clouds have been studied in many details \citep{Bally88, Oka98a, Oka01, Miyazaki00}. These surveys reveal a very complex medium whose nature, structure and kinematics is not fully understood yet. For instance, their formation and binding processes are still uncertain. 

More than 150 clouds have been identified but, because of projection effects, several clumps can be observed as one single cloud, so that the actual cloud number could be a bit higher \citep{Bally88}. \citet{Miyazaki00} reported 159 clouds at Galactic longitudes $-1^o< l < + 1^o42'$ and \citet{Oka01} detected 165 clouds in the region $-0^o.8 < l < 1^o.7$. However, several of them could belong to the 180-pc molecular torus and it is not clear how many really belong to the inner part. 
\citet{Launhardt02} estimated a volume filling factor of a few percents for this cold phase in the region $r<120$ pc, $|z|<50$ pc. Assuming the clouds have a 10 pc size, we derive a corresponding number of 135 clouds This region considered by \cite{Launhardt02} is smaller than the region we are interested in but again, several of these clouds could belong to the front part of the 180-pc molecular ring on the line of sight. Finally, we consider that at least 100 clouds belong to the inner CMZ.

The cloud size is fairly well constrained even if there is uncertainty in the cloud size definition. Clouds are observed to have radii between 1 pc and about 10 pc. Some clouds of a few tens of parsec size are also observed but they could be complexes of several smaller dense clouds. \citet{Miyazaki00} and \citet{Oka01} derived a mean radius of 3.7~pc and 6~pc respectively, which leads to use 5~pc as a typical radius.
However, clouds are thought to have very complex shapes and structures. They are often very extended with arc-, shell- or lane- shapes. Also, many large clouds are composed of several smaller clumps. Gas is observed to fill the space between the clumps with densities $<10^3$~cm$^{-3}$, lower than the mean density of the clumps ($10^4$ cm$^{-3}$). This gas is likely to give them a common dynamics. In that sense, the clouds may have a fractal structure similar to that of the standard ISM \citep{Falgarone91}.  

The cloud velocity is one of the less constrained parameters. Careful studies of the velocity distribution suggests that most of the molecular gas is in a torus at about 180~pc from the Galactic center \citep{Kaifu72, Scoville72, Binney91, Morris96}. Molecular clouds of about 10~pc size are thought to form in this torus, then detach and spiral inward with a significant velocity relative to the surrounding medium. 
The observed velocities are in a range between $V_{LSR}$=~ -250~ km/s and $V_{LSR}$=~ +250 km/s \citep{Bally88}. The clouds have a huge cloud-cloud velocity dispersion so that it is difficult to get a reliable rotation curve.  The cloud velocity dispersion within giant molecular complexes is about 30-50 km/s. This figure has been directly measured in various surveys and is in good agreement with the observed scale height of 50 pc of these complexes \citep{Oka98a}. As the hot gas is expected to have very smooth properties, this velocity dispersion is a minimal value for the cloud velocity relative to that gas, whatever the rotation profile is. 
Besides this dispersion, the rotation profile is not consistent with uniform circular rotation. \citet{Binney91} have worked out that the gas dynamics of this Central region is governed by a bar-potential. They suggested that the molecular ring at 180 pc is a transition region, corresponding to the Inner Lindblad Resonance of the bar: beyond this torus, the gas orbits along ellipses aligned with the bar axis (the so called $X_1$ orbits) and inside this torus, molecular clouds orbit on very elongated ellipses perpendicular to the bar axis (the $X_2$ orbits). 
Because of its high temperature, the 8 keV plasma is not likely to respond much to the bar potential. Moreover, if the pervasive magnetic field is generated by azimuthal currents, they must be localized in the molecular ring, providing the gas pressure gradient necessary to balance the magnetic pressure gradient. One can thus expect the magnetic field lines to rotate at the torus speed and drag the hot plasma with them. As a result, the relative velocity between the clouds and the surrounding material can be expected to be a substantial fraction of the cloud orbital speed (typically 200 km/s).
Furthermore many clouds are observed with a forbidden velocity, i.e. velocity of sign opposite to the rotation direction ($l<0$, $V_\mathrm{LSR} > 0$ and $l>0$,  $V_\mathrm{LSR} < 0$) and these velocities can be as high as 130 km/s  \citep{Oka98a}, which would imply a huge velocity relative to the ambient field.  
And last, the interpretation of filaments as resulting from the interaction of moving clouds with a pervasive magnetic field (as discussed below) requires a relative velocity of 50-150 km/s, typically 100~km/s \citep{Bally88, Sofue05}. Will will use 100 km s$^{-1}$ as the mean cloud velocity.

Many vertical filaments have been reported in the Galactic center region \citep{Yusef-Zadeh87c, LaRosa00a}. These filaments are non-thermal and magnetized, which indicates that the central region is magnetized too. The strength and topology of the magnetic field is a debated issue.
 On the one hand, it has been suggested that the filaments may trace a strong (B $\sim$ mG) vertical pervasive magnetic field \citep{Morris96}. Because in most cases they are associated with molecular clouds, these extended structures have been suggested to result from the interaction of the pervasive field with the clouds \citep{Benford88, Morris89, Lesch92, Rosso93, Serabyn94}. To drive vertical currents or guide accelerated particles along the filaments, the field has to be vertical. Its strength has been deduced from the observation that the field lines are not distorted by the cloud motion. This observation implies that this motion is sub-Alfv\'enic. Previous estimates  assumed that  the magnetic perturbations propagate in a medium of density comparable to that of the clouds (n $\sim10^{-3}$~cm$^{-3}$) and estimated mG strengths. However, if the Alfv\'en waves propagate in the faint hot diffuse plasma ($n \sim .1$ cm$^{-3}$), then, to keep the filaments straight over a height of 5 times the cloud's radius ($ v_\mathrm{A} > 5 v_\mathrm{c}$), this only requires: $B\gtrsim 100 \mu$G. 
A strong magnetic field is also invoked to confine the clouds. Indeed, the latter are not massive enough to be confined by gravity. They need an external pressure to balance the internal turbulent velocity. The hot plasma pressure is not strong enough to balance the ram pressure of the clouds but a pervasive field with $B\sim$~0.5~mG could do the job \citep{Miyazaki00}. 
On the other hand, it has also been suggested that the non thermal filaments may not trace a strong field but may rather correspond to local enhancements of a mean field in pressure equilibrium. The field could thus be lower than $B\sim$ 1 mG \citep{LaRosa00a}. Recent considerations on the diffuse non-thermal radio emission gave upper limits for the mean field. If these results are confirmed the field strength must be lower than $\sim 100 \mu$G \citep{LaRosa05}. In a hot helium plasma at 8~keV, the equipartition field is about $B \sim 100\mu$G. In this paper, we use this value as a reference but we also discuss the consequences of 10 $\mu$G-1 mG fields.

\section{The Braginskii viscosity}
\label{viscosity}
We review in this section the viscosity of magnetized plasmas, which is very different from the hydrodynamical one, usual in neutral gases, and needs to be discussed. 

\subsection{General properties}
Let us first consider a non magnetized plasma. Its viscosity is isotropic and it is thus governed by the sole coefficient :  
\begin{equation}
\eta_0 = .96 \frac{3\sqrt{m_i}}{4\sqrt{\pi} \lambda Z_i^4e^4} (k_BT)^{5/2}
\end{equation}
where $m_i$ and $eZ_i$ are the mass and charge of ions constitutive of the plasma \citep{Braginskii65}. We note that the temperature dependance is stronger than for neutral gases ($\propto T^{1/2}$) because the Coulomb cross section depends on the particle velocity whereas hard sphere-type collisions do not. This implies that a 8 keV helium\footnote{The viscosity of a hydrogen plasma would be 8 times higher.} plasma is highly viscous: 
\begin{equation}
\eta =  630 ~ \mbox{g~cm}^{-1}\mbox{s}^{-1} \left( \frac{k_BT}{8\mbox{ keV}} \right)^{5/2} \label{eta}
\end{equation}
The corresponding kinematic viscosity $\nu=\eta/\rho$ is then: $ \nu = 2.7\times10^{27}$~cm$^2$s$^{-1} $, tens of orders of magnitude higher than in usual fluids. 

As mentioned in sect. \ref{arena}, observations of the filaments perpendicular to the Galactic plane can be interpreted as tracers of a coherent, vertical magnetic field. Magnetic fields make the stress tensor anisotropic: the perpendicular mean free path is reduced to the Larmor radius, so that the diffusive properties are very different for the parallel and for the perpendicular direction. 
When applied to the transport of perpendicular momentum, it is found that the shear viscosity implying terms in $\partial_i v_j$ is reduced by the magnetic field whereas the bulk viscosity implying terms in $\partial_i v_i$ remains unchanged \citep{Braginskii65}. In a general manner, five different coefficients determine the stress tensor of a magnetized plasma. All of them can be expressed from $\eta_0$ as negative powers of $\Omega_\mathrm{c} \tau$ where $\Omega_\mathrm{c}$ is the cyclotron frequency and $\tau$ is the collision time. When the field is strong enough, all of them vanish but $\eta_0$. The collision time of the 8~keV helium plasma is \citep{Belmont05}:
\begin{equation}
\tau =  46\times10^3\mbox{ yr } \left(\frac{k_BT}{8\mbox{ keV}}\right)^{3/2} \left(\frac{n}{.035 \mbox{ cm}^{-3}}\right)^{-1} \label{taucoll}
\end{equation}
Even for the lowest estimates of the magnetic field strength ($B\sim1 \mu G$), the ratio of the highest coefficient to $\eta_0$ is:  $\eta_1/\eta_0\sim 10^{-20}$. The shear viscosity in the Galactic center region is fully inhibited by the magnetic field and only the bulk viscosity can dissipate energy.
The bulk viscous stress is \citep{Braginskii65}:
\begin{equation}
\vec{F} = \eta_0 \left( \frac{1}{3}\grad D - \partial_\parallel D ~\vec{e}_\parallel \right) \label{force}
\end{equation}
where the subscript $\parallel$ notes vector components along the field, $\vec{e}_\parallel$ is the unit vector pointing along the field direction, and 
\begin{equation}
D =  \dive \vec{v} - 3 \partial_\parallel v_\parallel
\end{equation}
is strongly related to the fluid compression. The associated power locally dissipated by the bulk viscosity is simply:
\begin{equation}
q = \frac{\eta_0}{3} D^2 \label{q}
\end{equation}

The implications of this viscosity are quite different from those of the shear viscosity. In particular, the Reynolds number may not be pertinent to characterize the viscous regime of a flow. For the typical cloud velocity $v_\mathrm{c}$ and radius $r_\mathrm{c}$, the Reynolds number would be:
\begin{eqnarray}
{\cal R}_e &=& 0.06~\left( \frac{r_\mathrm{c}}{5\mbox{ pc}} \right) \left( \frac{v_\mathrm{c}}{100\mbox{ km s}^{-1}} \right) 
\nonumber\\ 
& &~~~~~~~~~~~~~~~~~~
 \left( \frac{\rho}{2.~10^{25}\mbox{ g cm}^{-3}} \right) \left( \frac{k_BT}{8\mbox{ keV}} \right)^{-5/2} 
\end{eqnarray}
which would indicate an extremely viscous plasma.
However, this number was defined as the ratio of the inertial force over the viscous stress for a shear viscosity. The latter depends on terms in $\partial_i v_j$ which are of the order of  $v_\mathrm{c}/r_\mathrm{c}$ whereas the bulk viscosity implies terms as $\partial_i v_i$ which can be much smaller. As a consequence, the dimensionless number corresponding to the bulk viscosity, ${\cal R}_B$, is larger than the Reynolds number. For the cloud motion in the Galactic center, we can only ensure that ${\cal R}_B > .06$ and as a result, the flow around molecular clouds must be studied in details to be able to determine its viscous regime. 

\subsection{First example}
To better characterize the way the bulk viscosity acts on the flow, we perform here a 2-D analysis.
We study the flow around an infinite cylinder whose axis is parallel with the ambient magnetic field.
For a 2-D flow invariant along the field direction, $\partial_\parallel v_\parallel = 0$ and $D$ reduces exactly to the flow divergence: $D=\vec{\nabla}_\perp. \vec{v}_\perp$. A solution for the flow is presented as online material in appendix \ref{2-D};
and it is found that, in this geometry, the magnetic field only acts as an additional pressure. The solution is thus the same as for a neutral gas, except that the wave propagation speed is the transverse fast magnetosonic velocity $v_\mathrm{F}$. In that sense, the flow can be regarded as only resulting from standing and propagating fast perturbations. In the low-$\beta$ limit ($B\gtrsim100\mu$G), the dispersion equation for fast waves is: $\omega^2=k^2v_\mathrm{A}^2$. The cloud only excites waves with $\omega \sim v_\mathrm{c}/r_\mathrm{c}$ and $k\sim 1/r_\mathrm{c}$. The Alfv\'en Mach number in the hot and faint phase is:
\begin{eqnarray}
m_\mathrm{A} &=& 0.17 \left(\frac{\rho}{10^{-25}\mbox{gcm}^{-3}}\right)^{1/2} \left(\frac{B}{.1\mbox{mG}}\right)^{-1}
 \nonumber 
\left(\frac{v_\mathrm{c}}{100\mbox{kms}^{-1}}\right)
\end{eqnarray}
so that the clouds of the Galactic center region are very sub-Alfv\'enic. The cloud motion thus does not efficiently excite fast waves and we find that the flow around the cylinder is mostly incompressible. The first compressible contribution only appears to the order $(v_\mathrm{c}/v_\mathrm{F})^2$:
\begin{equation}
\dive \vec{v} \sim v_\mathrm{F}^2/r_vv_\mathrm{c}
\end{equation}
This leads to define a 2-D bulk Reynolds number:
\begin{equation}
{\cal R}_{2D} = 3\frac{v_\mathrm{c} r_\mathrm{c}}{\nu} \left( \frac{v_\mathrm{c}}{v_\mathrm{F}}\right)^{-2} 
\end{equation}
We see that the stronger the magnetic field, the higher the bulk Reynolds number. It has already been noted that the magnetic field inhibits the shear viscosity; we see here, for this 2-D case, that it also lowers the efficiency of the bulk viscosity.  Typically, the bulk viscosity is reduced by 2 orders of magnitude. For the properties of the Galactic center gas, the 2-D bulk Reynolds number is:
\begin{eqnarray} 
{\cal R}_{2D} &=& 5 ~\left(\frac{r_\mathrm{c}}{5\mbox{ pc}} \right) \left(\frac{v_\mathrm{c}}{100\mbox{ km s}^{-1}} \right)^{-1} \nonumber \\ & & ~~~~~~~~~~~~~~~~~~~~~~~ \left(\frac{B}{.1\mbox{ mG}} \right)^2 \left(\frac{k_BT}{8 \mbox{ keV}} \right)^{-5/2}
\end{eqnarray}

The 3-D situation is much more complex and it is not obvious how the 2-D conclusions can be adapted. First, in a fully three dimensional problem, the viscosity stress depends on the parallel motion of the flow. If $\partial_\parallel v_\parallel << \vec{\nabla}_\perp.\vec{v}_\perp$, the 2-D analysis holds and the above dimensionless number can be used to determine the viscous regime. On the contrary, if $\partial_\parallel v_\parallel \sim v_\mathrm{c}/r_\mathrm{c}$, the real Reynolds number corresponds to the bulk Reynolds number. And in between, when $\partial_\parallel v_\parallel \sim \vec{\nabla}_\perp.\vec{v}_\perp$, the bulk Reynolds number is intermediate. Thus the regime can be highly or weakly viscous depending on the exact flow properties. 
Second, whereas in this 2-D example only the fast waves seems to play a role, other modes, namely the Alfv\'en and slow ones make the flow properties in 3-D very different from this simple example.
As a consequence, a precise description of the 3-D flow around a spherical cloud is required to estimate the power dissipated in the plasma.

\section{Friction}
\label{heating}
In this section, we investigate how the viscosity can dissipate the energy of the molecular clouds in the plasma. For the sake of simplicity, we assume that the viscosity is not strong enough to modify the flow. We can first characterise the flow around one single cloud for an inviscid plasma. The dissipation is then estimated by using this inviscid solution and the viscosity properties.

\subsection{The clouds wake}
In this section, we refer to already existing literature to characterise the wake of molecular clouds flowing in the Galactic center region.

It is often assumed that strong currents can exist at least at the cloud surface. If the non-thermal filaments result from the field-cloud interaction, the clouds must be conducting, at least at their surface. Several processes can ionize the clouds enough to achieve a high conduction, such as the irradiation by local sources \citep[stellar clusters for example,][]{Morris96,Morris96b} or the Alfv\'en critical velocity ionization effect \citep{Galeev86,Morris89}. In the following, we will assume that they are perfect conductors. We are thus interested in the wake of a conductor moving in a magnetized plasma.

\begin{figure}
\resizebox{\hsize}{!}{\includegraphics{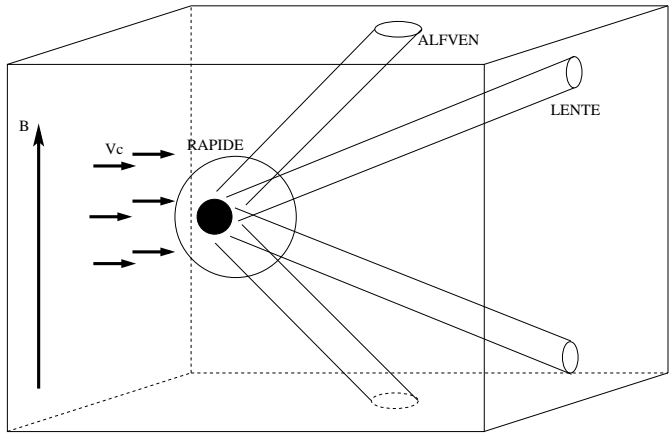}}
\caption{Schematic view of the MHD wake a a spherical cloud.} 
\label{wake}
\end{figure}

The behavior of electrically conducting bodies embedded in a flowing plasma has already been investigated but in different astrophysical conditions. \citet{Drell65} were among the first to address this question. Their goal was to explain the anomalous drag experienced by the Echo I artificial satellite in the earth magnetosphere. They introduced the important concept of \lq Alfv\'en wings\rq, which are standing Alfv\'en waves attached to the conductor.  
Many studies were also dedicated to the Alfv\'enic wake of the Jovian satellite Io, which is thought to have a good electrical conductivity \citep{Goldreich69}. An exact solution was found for the wings \citep{Neubauer80, Southwood80, Neubauer98}, which gives predictions for flow and field perturbations in excellent agreement with observations from Voyager 1 \citep{Acuna81,Belcher81,Barnett86}. Also, the possible contribution to the wake from non Alfv\'enic perturbations was early suggested \citep{Chu66,Wolf87,Linker88}. By considering conservation laws, \citet{Wright90} have shown that indeed the Alfv\'en wings cannot exist on their own: other disturbances must exist. In the MHD approximation three different modes can propagate in the plasma: the Alfv\'en mode but also the slow and fast magnetosonic ones. The motion of a moving body excites a priori the three modes so that the wake can be interpreted as constituted by three components. When the only source of disturbance is assumed to be the body itself, all the waves excited in its vicinity propagate away. As their propagation properties are different (see below), they can spatially separate and the three components can be easily identified at large distance. The general picture of such a wake is presented in fig. \ref{wake}. 
Numerical simulations present results very similar to this simple sketch. They confirm in particular the main properties of the Alfv\'en wings and show evidence for the slow and fast magnetosonic perturbations in the wake \citep{Linker88, Linker91,Linker98}. However, because of their higher complexity compared to the Alfv\'enic ones, the non Alfv\'enic disturbances were very little studied.

Since the three MHD modes have different properties, they are not subject to the bulk viscosity with the same efficiency (see sect. \ref{efficiency}). It is thus important to estimate their respective amplitude. The coefficients which determine the relative contribution of the three modes to the wake depend on the exact conditions at the cloud surface, such as the cloud speed, its shape, its conductivity, the magnetic flux its carries etc... The characteristics of the cold molecular clouds in the Galactic center region are different from those of Io and the artificial satellites and they may not excite the three modes in the same manner as satellites do. 

The relative velocities in the Galactic center are in particular very different from those of the satellites wake previously studied in the literature. 
The motion of satellites like Io or Echo I is rather fast compared to the wave speed: $v_\mathrm{c}\sim c_\mathrm{s} \sim v_\mathrm{A}$, whereas we are interested in very subsonic motions: $v_\mathrm{c} < c_\mathrm{s} \lesssim v_\mathrm{A}$. For this reason, our conclusions cannot be compared directly to the results of numerical simulations that exhibits solutions for low Mach numbers:  $m_\mathrm{A}=.5$ and $m_\mathrm{s}=.5-1$  \citep{Linker88, Linker91, Linker98}.

In the following, we will extensively refer to the solution proposed by \citet{Neubauer80}, which describes the wake of a magnetized satellite carrying the same magnetic field as in the surrounding medium. The magnetization of molecular clouds is not well understood. Their conductivity can be expected to be high enough to significantly prevent the magnetic field to diffuse in or out of the cloud on the crossing time of the cloud so that the magnetic flux threading the cloud remains constant on this typical time scale. The present magnetization thus depends on the poorly constrained initial magnetization and on the cloud history on long time scales. It is often argued that the field inside the clouds is made of external field lines which are tangled by the turbulent velocity of the inner clumps. The clouds magnetization must thus be a finite fraction of the external one. On the one hand, the cloud ionisation may be only partial so that the diffusion time may be shorter than the cloud life time. In such case, the magnetic field lines may have diffused on long time scales as the cloud moves, so that internal and external fields could now be comparable. On the other hand, it has been proposed that the external magnetic pressure could be responsible for the clouds confinement, pointing to a weaker cloud magnetization. 
The case where the cloud does not carry any field is very complex and has not been studied in details. Nevertheless, it has been proposed that this case would lead to an exact solution different from the wake with a strong magnetization, but with similar general properties \citep[][sect. 5.1.2]{Neubauer98}. For the sake of simplicity, we will thus assume that the clouds carry fields lines with comparable strength to those of the external medium.

In the next three subsections, we describe the properties  of the different modes and their respective amplitude for the subsonic motion of a conducting body in a strongly magnetised plasma ($\beta<<1$). We stress in particular the compression and the parallel velocity which govern the dissipation efficiency.
From now on, we refer to the vertical direction as the field aligned one and the horizontal plane as the plane perpendicular to $B$ and including the cloud-field relative velocity. The subscript $\parallel$ refers to the direction parallel to magnetic field, whereas the subscript $\perp$ refers to the components perpendicular to the mean field. We use the frame moving with the cloud. 

\subsubsection{The Alfv\'en wings}
The Alfv\'enic component of the wake has been most studied. Since the MHD equation admit a non-linear solution for the propagation of Alfv\'en waves, an exact non-linear solution has even been presented by \citet{Neubauer80}. 

The Alfv\'en waves are excited at the cloud surface by the transverse velocity and the conditions on the electric field; namely, the electric field tangent to the cloud surface has to vanish in the comoving frame. Their propagation is strictly along the field lines, which act as a guide. Thus their amplitude does not decay by geometric effects, as it would do if they were emitted isotropically.
As the Alfv\'enic perturbations travel vertically, the flow drags them horizontally. As a result, the Alfv\'en waves excited in the cloud vicinity are always located along the Alfv\'en characteristic, forming the so called \lq Alfv\'en wing\rq. There are two of them, one on each side of the cloud, corresponding to the two possible directions along the field. Each wing axis corresponds to an invariant direction. They are inclined at an angle $\alpha_\mathrm{A} = \tan^{-1}m_\mathrm{A}$ from the vertical direction. The clouds are very sub-alfv\'enic, so that the Alf\'en wings are almost vertical: $\alpha_\mathrm{A} \approx .02-0.2$ for B~$\approx$~.1-1~mG. 

By means of the Alfv\'en waves, the condition $E=0$ at the conductor surface propagates along the field lines, so that the comoving electric field vanishes in the whole flux tube threading the cloud, forming a cylindrical wing. Inside the wing, the magnetic field is aligned along its axis and the perpendicular velocity vanishes. The cloud thus entrains the whole flux tube and its full content along with it. As mentioned earlier, we assume in first approximation that the field threading the cloud is comparable with the external field. The wing size is thus of the order of the cloud diameter. Outside the wing, the stream and external field lines avoid the cylinder: this corresponds to perturbations decaying  as $1/r^2$ where $r$ is the distance to the wing axis. The surface of the cylinder is characterized by a current sheet which  closes the currents generated within the conducting cloud. In the solution given by \citet{Neubauer80}, each wing corresponds to a Poynting flux of: 
\begin{equation}
F_\mathrm{A} = \pi r_\mathrm{c}^2 \rho v_\mathrm{c}^2 v_\mathrm{A}
\end{equation}

As the stream avoids the whole flux tube, the perturbed velocity is comparable with the cloud velocity: $\delta v_\perp /v_\mathrm{c} \sim 1$ inside and in the vicinity of the cylinder. This wing is fully incompressible: $\delta \rho/\rho =0$, even at the non-linear stage. The linear solution has no parallel velocity, so that the linear Alfv\'en wing has $D=0$ and no dissipation. However, the non-linear solution allows a small parallel velocity which allows a finite $D$ and thus a small dissipation. When expanded to the first non-linear order, it is found that:
\begin{equation} 
v_\parallel /v_\mathrm{c} \sim m_\mathrm{A} \label{NL}
\end{equation}

As we will see, the Alfv\'en wing derived by Neubauer is the main contribution to the wake, but it does not fully satisfy the boundary conditions at the cloud surface, so that contributions from the other modes are needed.

\subsubsection{The slow wings}
For instance, the parallel velocity found in the non-linear solution for the Alfv\'en wings induces a net outflow. As the cloud cannot provide this material, this mass flux must be balanced by other perturbations. The respective role of the fast and slow modes is not well understood. \citet{Wright90} suggested that the matter supply can be at least partially attributed to the slow perturbation. 

The general propagation of slow magnetosonic modes is more complex than that of Alfv\'en ones but, in the low-$\beta$ limit, the slow waves are also guided along the field lines. As Alv\'en perturbations do, they are thus able to form wings on both sides of the clouds. The slow modes however propagate with a velocity close to the sound speed, slower than the Alfv\'en one, so that the slow wing angle $\alpha_\mathrm{s} = \tan^{-1}m_\mathrm{s}$ is larger. For a typical sound speed of 1000~km~s$^{-1}$, $\alpha_\mathrm{s} = 0.1$. For mG fields, the slow wing separates from the Alfv\'en wing at an altitude $h\approx2r_\mathrm{c}/m_\mathrm{s}$, i.e. $\sim100$~pc. For lower fields, they separate even farther, but when the field is in equipartition or weaker, slow waves do not strictly propagate along the field lines, so that they do not form a truly infinite wing. As long as the field is not too weak, the global shape is however similar. Since we consider a 70 pc height region, the two wings are thus superimposed in the whole emission region for any field strength we consider in this paper. In the linear approach, we can however study them independently. In the following, we will assume that the parallel velocity of the slow wave balances the parallel velocity of the Alfv\'en wings.
The polarization relation of slow waves gives: 
\begin{equation}
v_\perp = - \beta \frac{k_\parallel k_\perp}{k^2} v_\parallel
\end{equation} 
The slow wings are almost vertical so that $k_\parallel/k_\perp \approx m_\mathrm{s} << 1$. As a result, the transverse perturbed velocity remains very small in comparison to the relative velocity and to the Alfv\'enic transverse velocity. 
From the slow mode properties, we also get the perturbed density: $\delta \rho /\rho \sim m_\mathrm{A} m_\mathrm{s}$. Since the motion of molecular clouds is subsonic and sub-Alfv\'enic, the compression is very weak: $\delta \rho/\rho << 1$.
In the low-$\beta$ limit where the slow waves form a wing, the quantity $D$ in the slow wings is dominated by the contribution of the parallel velocity: $D \approx -2 k_\parallel v_\parallel$.

\subsubsection{The fast perturbation}
It is likely that fast perturbations are also excited.
It has already been mentioned that the Alfv\'enic perturbations in the solution from \citet{Neubauer80} extend beyond the cylinder frozen in the cloud. Nevertheless, the Alfv\'enic perturbations directly excited by the cloud surface can only propagate in the flux tube threading the cloud and Alfv\'enic perturbations propagating on external field lines cannot be excited by the cloud itself. They must be non-linearly excited by an other mode which would directly be generated by the cloud. The slow mode is guided by the field lines and can not contribute to this excitation, so that only a fast perturbation extending around the cloud can be responsible for the external structure of the Alfv\'en wing. This shows that the linear analysis must be handled with care, but also that there must be a fast component in the wake.

The general propagation of fast magnetosonic modes is complicated too but, in the low-$\beta$ limit, the fast waves propagate isotropically, as simple sound waves at the Alfv\'en speed. They thus cannot form a wing and their amplitude decays with the distance to the source. As a result, the fast perturbations excited by the cloud remain localized in its vicinity, just as in the classical hydrodynamical wake.

Because of the spherical propagation of fast waves, it is not easy to derive an estimate for $D$ in a fully 3-D case. We can however try to rely on the 2-D results. We have found that, for a subsonic and sub-Alfv\'enic motion, the cloud cannot excite waves that satisfy the dispersion equation of fast modes, so that the compression is very weak: $\delta \rho /\rho \sim m_\mathrm{A}^2$. We moreover note that this is also the case for a general 3-D wake in a non magnetised gas. The same conclusions must apply in 3-D for our MHD flow. We will thus use this relation for our following estimates.
In the low-$\beta$ limit, the polarization relations of the fast plane waves give: $k_\parallel v_\parallel = \beta k_\perp v_\perp$. Assuming it is still the case for spherical waves, we find that the parallel velocity has only a very small contribution in the quantity $D$, so that it is dominated by the perpendicular flow divergence: $D\sim \vec{\nabla}_\perp.\vec{v}_\perp$. If not, this result can be used as a minimal estimate of the dissipation in the fast perturbation.\\
\subsection{Viscosity efficiency}
\label{efficiency}
Because of their different properties, viscosity acts differently on the 3 components. In the absence of any other cooling mechanism, the dissipation efficiency can be determined by comparing the power dissipated by viscosity and the X-ray emission in the whole central region:
\begin{equation}
L \approx 4\times10^{37} \mbox{ erg s}^{-1} \left( \frac{h}{70\mbox{ pc}} \right)
\end{equation}
Here we specify this efficiency for the three different wake components.

\subsubsection{The Alfv\'en wings}
As mentioned earlier, the Alfv\'en waves are incompressible, and have no parallel motion in their linear form. Hence, in the Alfv\'en wing, $D=0$ and there is no dissipation due to the bulk viscosity. To lowest order, the energy in the Alfv\'en wings thus leaves the central region without being damped nor heating the plasma. 
However, we find that the integrated Poynting flux for both wings of the $N_\mathrm{c}$ clouds in the Galactic center is typically one order of magnitude higher than the X-ray luminosity:
\begin{eqnarray}
F_\mathrm{A} &=& 2.\times10^{38~} \mbox{erg~s}^{-1} \left(\frac{N_\mathrm{c}}{100} \right) \left( \frac{r_\mathrm{c}}{5\mbox{ pc}} \right)^{2} \left( \frac{B}{.1\mbox{ mG}} \right) \label{fluxalfven}  \\ 
&& ~~~~~~~~~~~
\left( \frac{\rho}{2.10^{-25}\mbox{ g cm}^{-3}} \right)^{1/2}  \left( \frac{v_\mathrm{c}}{100\mbox{ km s}^{-1}} \right)^{2}   \nonumber
\end{eqnarray} 
It is thus natural to wonder if secondary effect could dissipate a small fraction of this large energy. For 100 $\mu$G fields, 10\% would be required whereas for mG fields, only 1\% is needed.  In general, the damping of Alfv\'en waves is not very efficient, but different processes can dissipate a small fraction, as for example small magnetic irregularities, density gradients... Here, we concentrate on two possible viscous mechanisms: the damping of the non-linear parallel velocity and that resulting from a small compressibility of Alfv\'en waves propagating along magnetic field lines which have a large scale curvature. 

As mentioned previously, \citet{Neubauer80} could exhibit an exact non-linear solution for the Alfv\'en wings. This solution, although it is incompressible, involves a small parallel velocity which allows dissipation by the bulk viscosity. From eq. \ref{NL} we derive $D\sim 3 m_\mathrm{A}^2 v_\mathrm{c}/r_\mathrm{c}$. As long as the viscosity does not dissipate the whole energy flux, we can estimate the energy dissipated along the wing up to an altitude of 70~pc by summing the local dissipation rate over this height. For 100 clouds, we get the dissipation:
\begin{eqnarray}
Q_\mathrm{A1} &=& 2.\times10^{37}~ \mbox{erg s}^{-1}~\left(\frac{N_\mathrm{c}}{100} \right) \left( \frac{h}{70\mbox{ pc}} \right) \left(\frac{k_BT}{8\mbox{ keV}} \right)^{5/2} \nonumber\\ 
 && 
\left( \frac{\rho}{2.10^{-25}\mbox{gcm}^{-3}} \right)^{2}  \left( \frac{v_\mathrm{c}}{100\mbox{km s}^{-1}} \right)^{6}   \left( \frac{B}{.1\mbox{mG}} \right)^{-4} 
\end{eqnarray} 

Since the non linear parallel velocity is proportional to $v_\mathrm{c}/v_\mathrm{A}$, the magnetic field strongly inhibits this non linear dissipation and for fields above equipartition, this power is less than the radiative cooling. However, we find that for fields lower or equal to 100 $\mu$G, this energy is comparable with the radiative losses and can therefore account for the plasma heating.

\begin{figure}
\begin{center}
\resizebox{5cm}{!}{\includegraphics{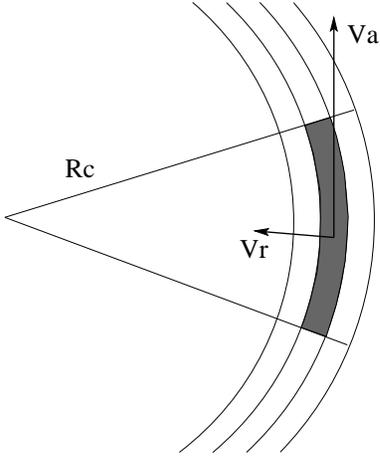}}
\end{center}
\caption{Illustration of the compression associated with the propagation of Alfv\'en-like perturbation in a curved magnetic field. The field lines are represented by the parallel curves and define a local Alfv\'en velocity $v_\mathrm{A}$, $R_\mathrm{c}$ is the curvature radius and $v_r$ is a transverse perturbation along the curvature direction. For straight lines, the Alfv\'enic transverse perturbation has no compression, whereas in a curved geometry, the volume of small elements of flux tubes changes with the distance to the curvature center: any displacement towards the curvature center drives a compression and any displacement away from the curvature center drives a depression.} 
\label{curve}
\end{figure}

On the other hand, the observed non-thermal filaments seem to be slightly bent, which may trace a large scale curvature. From observations, we estimate the radius of curvature being of the order of 100 pc. This is consistent with the hypothesis that the large-scale vertical magnetic field might be due to currents in the molecular torus at 150 pc.
In such conditions, the usual description of plane waves can not be applied exactly. In the limit where the radius of curvature is large in comparison with the wavelength, modes can be identified with the usual Alfv\'en, slow and fast ones, but with slightly different properties. For instance, Alfv\'en waves become weakly compressional: any Alfv\'enic displacement perpendicular to a curved mean field experiences a compression or a depression depending whether it points to or away from the curvature center. This is illustrated on fig. \ref{curve}.
This property can be interpreted as a coupling between the slow and Alfv\'en waves in a curved magnetic field \citep{Southwood85}. From fig. \ref{curve}, we can estimate the divergence of any Alfv\'enic perturbation to be of the order of:
\begin{equation}
\dive \vec{v} \approx \frac{v_r}{R_\mathrm{c}}
\end{equation}  
where $v_r$ is the perturbed velocity component projected along the curvature direction and $R_\mathrm{c}$ is the local radius of curvature. Since they are compressible, Alfv\'en waves in a curved magnetic field can be damped by the bulk viscosity, even at their linear stage. A more detailed treatment is presented in appendix \ref{curveB} (online material) where we show that the quantity $D$ responsible for the dissipation is actually:
\begin{equation}
D_\mathrm{A2} = \frac{2 v_\mathrm{c}}{R_\mathrm{c}} \left( 1-3\frac{c_\mathrm{s}^2}{v_\mathrm{A}^2} \right)
\end{equation}
In the low-$\beta$ limit, the corresponding dissipation for $N_\mathrm{c}$ clouds is:
\begin{eqnarray}
Q_\mathrm{A2} &=& 3.\times10^{37~} \mbox{erg~s}^{-1}~\left(\frac{N_\mathrm{c}}{100} \right) \left( \frac{h}{70\mbox{pc}} \right) \left(\frac{k_BT}{8\mbox{keV}} \right)^{5/2} \\ 
& &\left( \frac{v_\mathrm{c}}{100\mbox{km s}^{-1}} \right)^{2}    \left( \frac{r_\mathrm{c}}{5\mbox{pc}} \right)^2 \left( \frac{R_\mathrm{c}}{100\mbox{pc}} \right)^{-2} \nonumber
\left( 1-3 \frac{c_\mathrm{s}^2}{v_\mathrm{A}^2}\right)^2\label{fluxalfvenbis}
\end{eqnarray} 
We see that when $v_\mathrm{A}^2/c_\mathrm{s}^2 \sim 3$, i.e. when B $\approx$ 300 $\mu$G there is no dissipation. This is however a limit between to domains where the dissipation is significant. On the one hand, for stronger fields, the dissipation tends to a finite value which depends on the curvature radius. We find that a curvature radius $R_\mathrm{c}\lesssim 100$ pc is required for the damping to balance the radiative cooling in a mG field, which may be consistent with the observations. And on the other hand, for weaker fields, the dissipation is quickly comparable with the radiative losses and can become orders of magnitude stronger if the field strength is smaller than 100 $\mu$G. Obviously, the bulk viscosity cannot dissipate more that the Poynting flux. When our estimate exceeds the energy flux of the Alfv\'en wings, the viscosity starts changing the flow properties and our estimate fails. A full treatment of the viscosity is then required. However, it is likely to give a saturated solution where the Alfv\'en flux is fully dissipated and this energy is more than needed to balance the radiative cooling.

\subsubsection{The slow wings}
We have assumed the parallel velocity of the slow wing to be similar to that in the Alfv\'en wings. By the polarization relations of slow waves, this fully determine the slow wing amplitude and gives in particular $D\sim m_\mathrm{A} m_\mathrm{s} v_\mathrm{c}/r_\mathrm{c}$. When $\beta \lesssim 1$, this is larger than in the Alfv\'en wings because the slow wings are more inclined. To estimate the power dissipated in the whole central region, we can again sum the local dissipation rate over a height of 70~pc, and we find that the 100 clouds of the Galactic center can provide 
\begin{eqnarray}
Q_\mathrm{S} &=& 3 \times10^{36}~ \mbox{erg~s}^{-1} ~\left(\frac{N_\mathrm{c}}{100} \right)\left( \frac{h}{70\mbox{ pc}} \right)  \left(\frac{T}{8\mbox{ keV}} \right)^{5/2}  \nonumber\\ 
& & ~~~~~~~~~~~~
\left( \frac{\rho}{10^{-25}\mbox{ g cm}^{-3}} \right)  \nonumber \left( \frac{B}{.1\mbox{ mG}} \right)^{-2}  \nonumber \\ 
&& ~~~~~~~~~~\left( \frac{v_\mathrm{c}}{100\mbox{ km s}^{-1}} \right)^{6} \left( \frac{c_\mathrm{s}}{1000\mbox{ km s}^{-1}} \right)^{-2}  \label{slowflux}  
\end{eqnarray} 

We find that the power dissipated in the slow wing is one order of magnitude lower than the radiative losses. To reach a balance, weaker fields or a higher accretion rate are thus needed.

\subsubsection{The fast perturbation}
Fast waves are compressible. They could then be expected to be responsible for a strong dissipation. However, as we have seen, since the motion is subsonic, the compression remains weak whereas the velocity perturbation is large. If $\partial_\parallel v_\parallel \lesssim  \vec{\nabla}_\perp . \vec{v}_\perp$, the results found in the two-dimensional analysis are still valid in 3 dimensions and $D\sim m_\mathrm{A}^2 v_\mathrm{c}/r_\mathrm{c}$. This value is comparable with the non-linear Alv\'enic $D$. However, contrary to the latter, the fast perturbation is located in the cloud vicinity. The total efficiency is thus lower. We can estimate the total power dissipated by multiplying the 2-D results by the height of the cloud $2r_\mathrm{c}$. This leads to:
\begin{eqnarray}
Q_\mathrm{F} &=& 2.\times10^{35}~\mbox{erg~s}^{-1}~\left(\frac{N_\mathrm{c}}{100} \right) \left( \frac{h}{5\mbox{ pc}} \right) \left(\frac{k_BT}{8\mbox{ keV}} \right)^{5/2}  \nonumber\\ 
& & 
 \left( \frac{\rho}{10^{-25}\mbox{g cm}^{-3}} \right)^{2} \left( \frac{v_\mathrm{c}}{100\mbox{km s}^{-1}} \right)^{6}  \left( \frac{B}{.1\mbox{mG}} \right)^{-4} 
\end{eqnarray} 
This is two orders of magnitude too small to contribute significantly to the dissipation. As we have noted earlier, if the parallel velocity plays a more important role in the 3-D situation, then this might underestimate the dissipation. Moreover, even if the fast perturbation amplitude and the associated dissipation decay with distance to the cloud, the dissipation integrated over far spherical shells might remain significant. This would lead to using a larger scale height. It is thus possible that the total dissipation in the fast perturbation is stronger than the above estimate. A more detailed treatment is therefore required but the dissipation is unlikely to dominate over the Alfv\'enic dissipation, so that we do not discuss this possibility further on. 

\section{Discussion and conclusion}
\label{discuss}

\begin{figure}
\resizebox{\hsize}{!}{\includegraphics{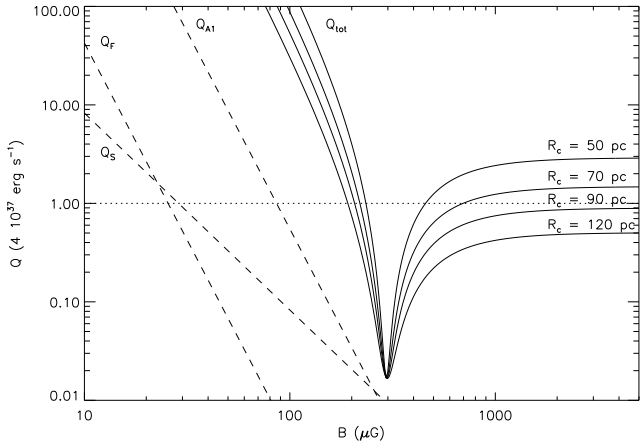}}
\caption{Total dissipation versus magnetic field strength for different radii of curvature $R_\mathrm{c}$. The power is normalized to the X-ray luminosity in the Galactic center region, so that 1 is the power required to balance the radiative cooling. Contributions $Q_{A1}$, $Q_{S}$ and $Q_\mathrm{F}$ for the power dissipated in the Alfv\'en wing by non-linear effect, in the slow wing, and in the fast perturbation respectively, are also plotted, although the estimates for the slow and fast perturbations lose accuracy below 100 $\mu$G.} 
\label{efficiencyplot}
\end{figure}

As a conclusion, we find that the dissipation is different in each component of the wake and that the global efficiency depends on the field strength. For all strengths, dissipation within the fast and slow wings is negligible and most of the dissipation occurs in the Alfv\'en wing. We find that the dissipation resulting from the curvature of field lines dominates for any field. Figure \ref{efficiencyplot} shows the total dissipation and contributions for each wing, as a function of the field strength.
Three domains can be observed:
\begin{itemize}
\item
For low fields: B $\lesssim$ 100$\mu$G, both the non-linear and the curvature damping dissipate enough power to balance the X-ray luminosity. The dissipation due to the curvature of field lines seems to be dominant, for any reasonable radius of curvature. In this strength range for the magnetic field, the dissipation in the Alfv\'en wings can be responsible for the plasma heating.
\item
For intermediate fields: B $\approx$ 0.2-0.5 mG, the dissipation is more than one order of magnitude lower than the losses by radiation.
\item
For strong fields: B $\gtrsim$ 1 mG, we find that none of the components can dissipate enough energy to balance the radiative cooling in a straight geometry. Nevertheless, a large scale curvature with $R_\mathrm{c}\lesssim$ 100 pc would be responsible for a sufficient dissipation.
\end{itemize}

We see that except for the indirect dissipation, the magnetic field strongly limits the efficiency, favoring low fields. However it must be kept in mind that, for too weak fields, some of our results are not valid anymore. Indeed, we used the low-$\beta$ plasma limit to allow a very simple characterization of the slow and fast perturbations. In this limit, the slow modes are guided and the propagation of fast sonic modes is isotropic. When $B\lesssim 100\mu$G, the field is below equipartition and the properties of the fast and slow waves are more complex. However, it is likely that as long as the field is not too weak (B $>$ 10 $\mu$G), our related conclusions remain valid in order of magnitude. The description of the Alfv\'en wing does not depend on any assumption on the field strength, so that even for weak fields, our related conclusions still hold.

Our model however strongly depends on the field geometry, which is debated too. The field has for example been suggested to be turbulent and mostly toroidal as in the rest of the Galaxy \citep{Tanuma99}. In such case, our results cannot be applied to describe the interaction of the cold molecular clouds with the magnetic field. Nevertheless, the viscosity may still play a role. In a vertical field, most of the energy can flow away as Alfv\'en waves without being dissipated. In a turbulent medium, it has to stay within the plane. Coupling with other modes finally generates compressible modes which then have to be damped by viscosity. For these reasons, it is likely that, even in a turbulent field, the motion of the molecular clouds can participate to the plasma heating. The dissipation in a turbulent medium requires a dedicated investigation and it is not excluded that it might be even stronger than in the straight geometry. 

These results also depend on the statistical properties of the clouds, which are poorly constrained.
In particular, we have assumed clouds of spherical shape, which is obviously not accurate for the clouds in the Galactic center. The dissipation generated by their extended shapes and fractal structure cannot be easily derived, but local enhancements of the fluid compression are expected so that the global heating could be stronger. Following the 2-D analysis presented in appendix \ref{2-D}, we computed numerically solutions for non-cylindrical clouds flowing in a viscous medium. We found that multi-pole shaped clouds may dissipate several times more than a simple spherical cloud. The real structure of the clouds is far more complex than multi-poles and could result in an even more efficient dissipation.

The dissipation rate in the non-linear Alfv\'en wings, the slow wings and the fast perturbation also strongly depends on the cloud velocity: $Q\propto v_\mathrm{c}^6$. Any uncertainty on the estimate of this parameter can therefore have a strong effect on the order of the dissipation. In particular, in our estimates, we used an average value for the velocity whereas we should integrate the power resulting from each cloud over the velocity distribution. Such an integration would favor the fastest clouds and give a higher total dissipation rate. As the cloud velocity dispersion is large, this effect may be important enough to make the dissipation in the slow wing competitive with the radiative losses, even for intermediate field strengths.

We also note that the accretion of clouds from the molecular ring at 180 pc is not thought to be continuous. Some clues seem to indicate that it could be strongly intermittent with catastrophic events every $\sim 10^7$ yr \citep{Stark04}. The number of clouds and their velocity are expected to be higher during these events, providing a stronger heating. Because the radiative cooling time is very long ($>10^8$ yr) compared to the inter-event time, such a past heating would not be distinguished from a present one.

A very interesting property of this heating mechanism is that the viscosity regime seems to correspond exactly to the maximal efficiency for dissipation. We have presented here a MHD analysis of the cloud-field interaction. This holds as far as the collision time is shorter than the typical time scale of the problem. From eq. \ref{eta}, it is seen that when the friction heats the plasma, the latter becomes more and more viscous. But meanwhile, the collision time increases (eq. \ref{taucoll}). The plasma thus reaches a temperature where the regime becomes collisionless. If the temperature keeps growing, the collisions become so rare that the results from \citet{Braginskii65} fail. The wave-particle interaction must then be studied in a kinetic formalism, leading to damping by magnetic pumping; but as the viscosity basically results from collisions, it must drop in the collisionless regime. The maximal dissipation is thus reached when the collision time is of the same order of magnitude as the typical time scale of the problem. In the case of the cloud motion, this time is the cloud crossing time: $\Delta t = r_\mathrm{c}/v_\mathrm{c} \approx 48\times10^3$~yr, showing that the conditions are optimal for a maximal dissipation. This naturally suggests a self-regulation mechanism where the viscous friction heats the plasma to the temperature that makes the regime weakly collisional. Then, the efficiency drops and the temperature saturates. Given the strong dependance of the dynamical viscosity and the collision time with temperature, this gives a precise saturation temperature, which could be $k_B$T=8~keV.

The overall interaction of the moving cloud with the ambient field exerts a drag on the cloud and must slow it down, causing it to accrete toward the Galactic center. In the processes we have presented here, most of the energy taken from the cloud goes away as Alfv\'en waves, of which only a fraction is dissipated by the viscous processes we have described. The rest of the energy leaves the X-ray emitting region without contributing to the heating but does participate to the drag on the cloud. This phenomenon successfully explained the unexpected drag on Echo I \citep{Drell65} and may be significant for the molecular clouds too.  We estimate this energy to be $2 \sim 10^{36}$~erg~s$^{-1}$ per cloud. This gives a typical drag-induced radial velocity of $v_r \sim 1$~km~s$^{-1} \left(B/1 \mbox{ mG}\right)$ and an accretion time of $\sim 10^8$~yr $\left(B/1 \mbox{ mG}\right)^{-1}$. For strong fields, these results are comparable with the other accretion mechanisms such as the dynamical friction with stars \citep{Stark91}. They are also consistent with the accretion rate needed to sustain a constant accretion from the external molecular torus at 180~pc and explain the mass profile \citep{Morris96}. The pervasive field in the Galactic center may thus significantly contribute to driving the gas accretion. 

As a conclusion, we find that in spite of the strong vertical field, the viscosity in the Galactic center region can dissipate enough of the kinetic energy of the molecular clouds that orbit in the CMZ to balance the X-ray luminosity. This process not only provides the necessary power, but is also able to bring the gas to the observed temperature of 8 keV, which turns out to be the one where this process is most efficient. It should also contribute to the drag causing the clouds to accrete at this intermediate scale between the large-scale, bar-dominated flow and the smaller scale of the central black hole vicinity.

\begin{acknowledgements}
The authors thank M. Morris, M. Muno, S. Cowley and T. Chust for numerous and very enriching discussions.
\end{acknowledgements}

\bibliographystyle{aa}
\bibliography{GCbib2} 

\appendix
\section{The wake of an infinite cylinder}
\label{2-D}

In this appendix, we study the two-dimensional subsonic, sub-Alfv\'enic MHD flow around a cylinder for any viscous regime. This allows us to proceed by expansion starting from an incompressible solution.

\subsection{Equations}

We study the flow around a vertical cylinder of radius $r_\mathrm{c}$. The magnetic field far from the cylinder is assumed to be parallel with the cylinder axis. The flow at infinity has a constant velocity $v_\mathrm{c}$ and is perpendicular to the axis.

In this simple two-dimensional case, the viscous stress and the Lorentz force write very easily. 
On the one hand, the viscous stress is proportional to the compression of the fluid:
\begin{equation}
\vec{F}_\eta = \frac{1}{3} \grad_\perp \left( \eta_0\vec{\nabla}_\perp . \vec{v}_\perp \right) 
\end{equation}
On the other hand, in this geometry, there can be no perpendicular perturbed magnetic field so that the field lines remain vertical. They only compress and avoid the cylinder. As a result, the magnetic field only acts on the plasma dynamics by its pressure that is proportional to the density.

Finally, the system constituted by the mass conservation and the equation of motion reads:
\begin{eqnarray*}
\dive \left( \rho \vec{v} \right) &=& 0 \\
\rho (\vec{v}.\grad) \vec{v} &=& -v_m^2 \grad \rho+ \frac{1}{3} \grad \left(\eta
\vec{\nabla} . \vec{v} \right) 
\end{eqnarray*}
where $v_m^2 = c_\mathrm{s}^2+v_\mathrm{A}^2$ is the fast magnetosonic velocity.
It is non linear and cannot be solved at once but further assumptions can be made. As noticed in the main body of the paper, the motion of the molecular clouds is subsonic and sub-Alfv\'enic. We can thus restrict our study to this case and order the equation with the small parameter:
\begin{equation}
\epsilon = \left( \frac{v_\mathrm{c}}{v_m^\infty} \right)^2 
\end{equation}
where $v_m^\infty$ is the fast magnetosonic at infinity upstream.
From now, we use dimensionless quantities. Velocities, density and viscosity are normalized by their value at infinity upstream whereas distances and gradients are normalized by the cylinder radius. Then, the equation of motion reads:
\begin{equation}
\epsilon \rho (\vec{v}.\grad) \vec{v} = -v_m^2\grad \rho + \frac{1}{{\cal R}_{2D}} \grad \left( \eta \dive \vec{v}  \right) 
\end{equation}
where we have defined the bulk Reynolds number:
\begin{equation} 
{\cal R}_{2D} = \frac{3r_\mathrm{c} v_\mathrm{c}}{\nu_\infty \epsilon} 
\end{equation}
We can now order these equation with respect to $\epsilon$. 

	\subsection{Lowest order: incompressible flow}
To lowest order, the inertial terms do not appear. By integrating once the stream aligned component of the Euler equation, we get:
\begin{equation}
h_0(\rho_0) +  \frac{\eta_0}{\eta_\infty{\cal R}_{2D}^0}  (\vec{v}_0.\grad)\rho_0 = \mbox{cste}  \label{Euler0}
\end{equation}
where $h_0=\frac{c_\mathrm{s}^2}{\gamma-1}+2v_\mathrm{A}^2$ is an extended enthalpy for the fluid. Equation \ref{Euler0} is a first order differential equation for $\rho_0$ and there is no solution that does not diverge but the incompressible one: $\rho_0 = 1$, $\dive \vec{v}_0=0$. 
In cylindrical coordinates $(r,\theta)$ and with regular boundary conditions ($v_r=0$ at the cloud surface and $v_\infty=v_c$) we find: 
\begin{eqnarray}
v_r^0 &=& \cos{\theta} \left( 1-\frac{1}{r^2} \right)\\
v_\theta^0 &=& -\sin{\theta} \left( 1+\frac{1}{r^2} \right) 
\end{eqnarray}
Figure \ref{plot} shows this solution where stream lines avoid the cylinder without compressing. This solution is purely incompressible, so that the bulk viscosity can not dissipate energy.

	\subsection{First order: compressible solution}
\begin{figure}
\resizebox{\hsize}{!}{\includegraphics{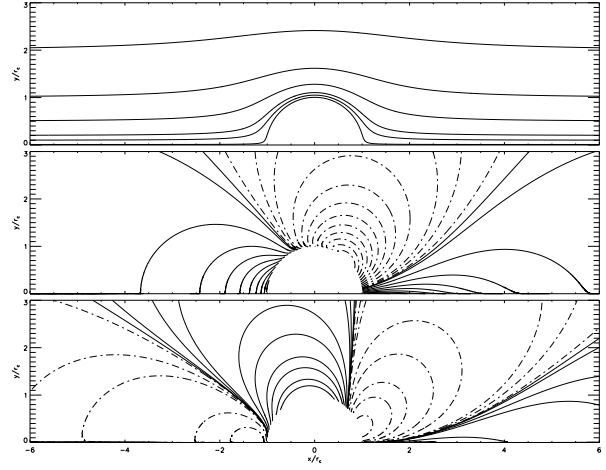}}
\caption{Total dissipation per unit height in the MHD flow around an infinite cylinder for ${\cal R}_{2D}$. Flow comes from the left hand side. The upper, mid and lower panels show the main velocity, the perturbed density and the perturbed flow divergence, respectively. Dashed and plain contours are negative and positive contours, respectively.} 
\label{plot}
\end{figure}

To following order, there are compressible terms. As before, we can integrate once the equation of motion to:
\begin{equation}
(\vec{v}_0.\grad) \rho_1 + {\cal R}_{2D}^0 \rho_1 = \frac{\epsilon{\cal R}_{2D}^0}{2} (1-v_0^2)  \label{euler1}
\end{equation}
Contrary to the zeroth order equation, the source vanishes far from the cylinder. We can thus find a non diverging solution. The perturbed density is then found by integrating this equation over the stream lines.   In the limit cases where ${\cal R}_{2D} << 1$ or ${\cal R}_{2D} >> 1$, the solution can easily be found whereas it is easier to make the integration numerically otherwise. Figure \ref{plot} gives the perturbed density $\rho_1$ for ${\cal R}_{2D} =1$. 
This solution is compressible so that, to this order, the bulk viscosity dissipates energy. The dissipation is proportional to the square of the flow divergence which is easily derived from the perturbed density by the mass equation: 
\begin{equation}
(\vec{v}_0.\grad) \rho_1  = -\dive \vec{v}_1  \label{masse1}
\end{equation}
Figure \ref{plot} shows the contours for the flow divergence.

\begin{figure}
\resizebox{\hsize}{!}{\includegraphics{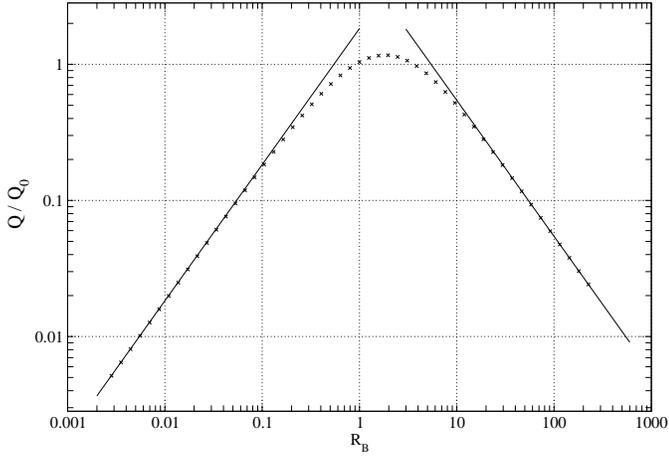}}
\caption{Dissipation rate as a function of the 2-D bulk Reynolds number. Here, $Q_0 = \rho_0 r_\mathrm{c} v_\mathrm{c}^5/v_m^2$. Points are the result of numerical integration of equation \ref{euler1}. The two asymptotes $Q/Q_0 = 1.83 {\cal R}_{2D}$ and $Q/Q_0 = 5.45{\cal R}_{2D}^{-1}$ result from its analytical integration in the viscous and non viscous regimes.} 
\label{puiss}
\end{figure}

To get the total dissipation per unit height, the local one must be integrated over the horizontal plane ($(x,y)$. In a general manner, it depends in the viscous regime and can be written
\begin{equation}
Q/ Q_0  = \frac{1}{{\cal R}_{2D}} \int \left(\frac{\dive \vec{v}_1}{\epsilon}\right)^2 dxdy
\end{equation}
where
\begin{equation}
Q_0 = \rho_0 r_\mathrm{c} \frac{v_\mathrm{c}^5}{v_m^2}
\end{equation}
is the typical dissipation and all the quantities in the right hand side are dimensionless. For comparison, the dissipation by the usual shear viscosity would be: 
\begin{equation}
Q_{\rm shear} \sim \rho_0 r_\mathrm{c} v_\mathrm{c}^3
\end{equation}
If the viscosity coefficient were equal, the bulk viscosity would be lower than that by the shear viscosity, by a factor $\epsilon$.
Figure \ref{puiss} shows the dissipation by the bulk viscosity as a function of the bulk Reynolds number.

\subsection{Conclusion}
We find that the flow in the subsonic limit can be decomposed as a main incompressible solution and a small compressible perturbation. 
The dissipation rate depends on the viscous regime. 
The maximal dissipation occurs for the intermediate regime when ${\cal R}_{2D} \sim 1$. The corresponding estimate is $\epsilon=(v_\mathrm{c}/v_m)^2$ times smaller than the dissipation which would occur via shear viscosity without magnetic field.

\section{Propagation of Alfv\'enic perturbations in a curved magnetic field}
\label{curveB}

In this appendix, we derive the characteristics of the Alfv\'en waves propagating in a curved magnetic field. As soon as the medium is inhomogeneous, the plane waves are not the eigensolutions and the true solutions have often a complex behavior. In the limit where the irregularities are weak, the solutions are similar to plane waves, but with slightly different properties.  In particular, we can expect the Alfv\'en waves to gain a compressional component, and thus to become subject to viscous damping. Here we write the full waves equations in a curved equilibrium. We then expand the equations by considering the curvature of field lines as very weak on the wavelength scale. This expansion gives the modified waves properties. 

\begin{figure}
\resizebox{\hsize}{!}{\includegraphics{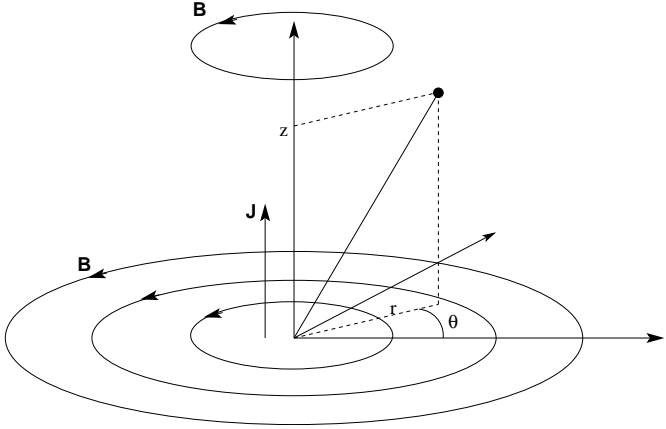}}
\caption{The equilibrium magnetic field and notation definitions} 
\label{cylgeom}
\end{figure}

To model the effect of field curvature, it is more convenient to use a cylindrical symmetry with invariance along the vertical axis $z$ and the azimuthal angle $\theta$. As shown in fig. \ref{cylgeom}, the equilibrium magnetic field is assumed to be purely azimuthal, so that the corresponding current is in the vertical direction:
\begin{equation}
\vec{B}_0 = B_0(r) \vec{e}_\theta ~~~~~~~~~~ \vec{J}_0 = \frac{1}{r} \partial_r (rB_0) \vec{e}_z
\end{equation}
In a static equilibrium, the pressure gradient needs to balance the Lorentz force:
\begin{equation}
\partial_r P_0 + \frac{1}{r^2}\partial_r \left(  \frac{r^2 B_0^2}{2} \right) = 0
\end{equation}
We note that in this geometry, a current-free equilibrium is obtained for $B_0 \propto 1/r$. Nevertheless, for reasons that will be explained later, we will use a specific equilibrium that makes $\omega_\mathrm{A}^2=v_\mathrm{A}^2/r^2=B_0^2/4\pi\rho_0r^2$ constant. 

\subsection{General perturbed equations}
We shall now perturb the system with small quantities $\rho_1$, $\vec{v}_1$ and $\vec{b}_1$.  We perform a Fourier series in $\theta$ and a Fourier transform in $z$, but, since the radial direction is not homogeneous, we keep the radial dependancy:
\begin{eqnarray}
\vec{B}_{tot}&=&\vec{B}_0 +\vec{b}_1(r)e^{-i(\omega t-m\theta-k_z z)}  \\
\vec{V}_{tot} &=& \vec{0} + \vec{v}_1(r)e^{-i(\omega t-m\theta-k_z z)} \\
\rho_{tot} &=& \rho_0 + \rho_1(r)e^{-i(\omega t-m\theta-k_z z)}
\end{eqnarray}
From now, we drop the subscript and denote by \rq ~ the radial derivative. We express the whole wave system as function of the three perturbed velocity components $v_r$, $v_\theta$, and $v_z$.
With these notations, the mass and induction equations read respectively:
\begin{eqnarray}
i\omega \rho/\rho_0 &=& v_r'  + \frac{im}{r} v_\theta + i k_z v_z + \frac{v_r}{r}\left( 1 + r\frac{\rho'_0}{\rho_0} \right) \label{masse} \\
i\omega \vec{b}/B_0 &=& -\frac{im}{r} \vec{v} \\ 
& +& 
\left( v_r' + \frac{im}{r} v_\theta + ik_z v_z + \frac{v_r}{r} \left( -1 + \frac{(rB_0)'}{rB_0}   \right) \right) \vec{e}_\theta \nonumber
\end{eqnarray}
The largest difficulty in this geometry is the radial dependancy that may not be simply expressed as a Fourier coefficient. To deal at best with this, it is easier to express the system as a function of the radial velocity $v_r$. After substituting $\rho$ and $\vec{b}$ in the vertical and orthoradial components of the equation of motion, the azimuthal and vertical perturbed velocities can be formulated:
\begin{eqnarray}
v_\theta &=& -\frac{im}{r{\cal D}}c_\mathrm{s}^2 \left\{ \left( \omega^2 - \frac{m^2}{r^2} v_\mathrm{A}^2\right)v_r' \right. \nonumber\\ 
& & ~~~~~~~~~~~~~~~~+ 
\left. \left(\omega^2-\frac{m^2}{r^2}v_\mathrm{A}^2-2k_z^2v_\mathrm{A}^2 \right)\frac{v_r}{r} \right\} \label{vtheta}\\
v_z &=& \frac{ik_z}{{\cal D}} \left\{ \left(\frac{m^2}{r^2}c_\mathrm{s}^2v_\mathrm{A}^2 -\omega^2v_\mathrm{F}^2\right)v_r'  \right. \nonumber\\ 
& & ~~~~~~~~~~~~~~
+ \left. \left( \omega^2(v_\mathrm{A}^2-c_\mathrm{s}^2)-\frac{m^2}{r^2}c_\mathrm{s}^2v_\mathrm{A}^2 \right) \frac{v_r}{r}\right\} \label{vz}
\end{eqnarray}
with 
\begin{equation}
{\cal D} = \omega^4 - v_\mathrm{F}^2\left(\frac{m^2}{r^2}+k_z^2\right) \omega^2 + \frac{m^2}{r^2}c_\mathrm{s}^2v_\mathrm{A}^2\left( \frac{m^2}{r^2}+k_z^2\right)
\end{equation}
where we have used the fast magnetosonic speed $v_\mathrm{F}^2 = v_\mathrm{A}^2+c_\mathrm{s}^2$.

To get the dispersion equation, these velocities must be substituted in the radial component of the equation of motion. It can eventually be written in the following manner:
\begin{equation}
\frac{\partial}{\partial r} \left[\frac{1}{r} P(\omega,m,k_z,r) \frac{\partial}{\partial r}rv_r\right] + Q(\omega,m,k_z,r) v_r =0 \label{HL}
\end{equation}
where:
\begin{eqnarray}
P &=& \frac{\rho_0}{{\cal D}} v_\mathrm{F}^2 \left( \omega^2-\frac{m^2}{r^2}v_\mathrm{A}^2\right)\left( \omega^2 -\frac{m^2}{r^2}\frac{c_\mathrm{s}^2v_\mathrm{A}^2}{v_\mathrm{F}^2}\right) \label{polyP}\\
Q &=& \rho_0\left(\omega^2-\frac{m^2}{r^2}v_\mathrm{A}^2 \right) - \frac{4k_z^2\rho v_\mathrm{A}^4  }{r^2 {\cal D}} \left( \omega^2 - \frac{m^2}{r^2} c_\mathrm{s}^2 \right)  \nonumber\\ 
& & ~~~~+ 
r\left\{ \frac{B_0^2}{r^2} - \frac{2k_z^2B_0}{r^2{\cal D}} \left( \omega^2 v_\mathrm{F}^2- \frac{m^2}{r^2} c_\mathrm{s}^2 v_\mathrm{A}^2 \right) \right\} ' \label{polyQ}
\end{eqnarray}
This equation is the generalized Hain-L\"ust equation \citep{HL58,Goedbloed71,Goedbloed04}. In a general manner, $P$ vanishes at specific radii. Solving this equation with boundary conditions thus leads to deal with singularities. This effect is not due to curvature but to the gradients of Alfv\'en velocity and of the parallel derivative $k_\parallel = \frac{im}{r}$, which are not of interest here. Here, we concentrate on the curvature effect. To this purpose, we refer to the equilibrium quoted earlier that makes $\omega_\mathrm{A}^2=k_\parallel^2 v_\mathrm{A}^2$ constant. In this case, the Alfv\'en condition $\omega^2-\omega_\mathrm{A}^2=0$ can be satisfied for any radius simultaneously and the singularity disappears.

\subsection{Modified propagation properties}
When $\omega^2 \rightarrow \omega_\mathrm{A}^2$, $P$ vanishes but $Q$ does not. As a result, pure Alfv\'en waves are not solutions of the full dispersion equation \ref{HL}. However, we are only interested in a large scale curvature of the field lines. We thus only consider perturbations whose typical scale length is shorter than the radius of curvature.  In this limit, the difference between the exact Alfv\'en-like solutions and the plane Alfv\'en waves must be small. 
Hence we now expand the equations by assuming that the perturbed system varies faster than the background quantities, i.e. on length scales smaller than $r$. As we are only interested in Alfv\'en waves whose behavior is mostly determined by their parallel wavelength, we only assume that the parallel wavelength is shorter than $r$: 
\begin{equation}
m >> 1
\end{equation}
The expansion is thus done with the small parameter $1/m^2$. The lowest order describes the plane propagation of the Alfv\'en waves and higher orders introduce curvature effects. $P$ is a lowest order term: $P=P_{-1}\sim m^2$ and $Q$ contains both lowest and first terms: $Q=Q_{-1}+Q_0$ with
\begin{equation}
Q_{-1} = \rho_0\left(\omega^2-\frac{m^2}{r^2}v_\mathrm{A}^2 \right)
\end{equation}
The following order term $Q_0$ introduce the curvature effects and is responsible for a shift in the Alfv\'en frequency. To find this shift, we can write $\omega = \omega_{-1} + \omega_0$, substitute in eq. \ref{HL} and identify order per order. To lowest order, we find the propagation relation of plane waves: 
\begin{equation}
\omega_{-1}^2= \frac{m^2}{r^2} v_\mathrm{A}^2
\end{equation}
and to first order, we find:
\begin{equation}
\omega_0\omega_{-1} = \frac{1}{r} \frac{(Pr^{2\gamma})'}{\rho_0 r^{2\gamma}} \left( 1 - \frac{(\rho_0(rv_r)')'}{k_z^2 \rho_0 v_r}\right)^{-1}
\end{equation}
We can furthermore assume that the radial gradient for the perturbed velocity is also stronger than for the background profiles. In a WKB approximation, the radial derivative can be written $\partial_r v_r \approx ik_r v_r$ and: 
\begin{equation}
\omega_0\omega_{-1} = \frac{1}{r} \frac{(P_0r^{2\gamma})'}{\rho_0 r^{2\gamma}} \left( 1 + \frac{k_r^2}{k_z^2}\right)^{-1}
\end{equation}
The shift in the Alfv\'en frequency depends on the background profiles. We will however see in the next subsection that this shift is not important for our concern.

\subsection{Modified compression and viscosity}
The general compression of the perturbed flow can be derived from eq. \ref{vtheta} and \ref{vz}:
\begin{eqnarray}
\dive \vec{v} &=& \frac{\omega^2}{\cal D}\left( \omega^2-\frac{m^2}{r^2} v_\mathrm{A}^2\right) \frac{1}{r} (rv_r)' -2 \frac{\omega^2 k_z^2 v_\mathrm{A}^2}{\cal D} \frac{v_r}{r} 
\end{eqnarray}
The divergence involves two kinds of terms. Those in $v_r'$ corresponds to the usual compression of sonic modes, also existing in a straight configuration. For plane Alfv\'en modes, these terms vanish. The other terms, in $v_r/r$, result from the field curvature. 
By substituting $\omega=\omega_{-1}+\omega_0$ in this equation, we find:
\begin{equation}
\dive{v} = 2\frac{v_r}{r} \left( 1 -S \right)
\end{equation}
where $S$ is a term directly resulting from the frequency shift:
\begin{eqnarray}
S &=& \frac{(P_0 r^{2\gamma})'}{r^{2\gamma}B_0^2} \frac{(rv_r)'/r}{k_z^2v_r - (\rho_0 (rv_r)')' /\rho_0} \\
 &\approx &  \frac{ik_r}{k_z^2 +k_r^2} \frac{(P_0 r^{2\gamma})'}{r^{2\gamma}B_0^2}
\end{eqnarray}
The second line has been written for $k_r >> 1/r$. Because of $S$, the general compression depends on the background profiles. However, it can be noted that in the above limit ($k_r r>> 1$), the frequency shift is very small and $S \sim 1/(k_r r) <<1$. 
Eventually, when neglecting the variation of the background quantities on wavelength scale, but keeping the curvature effects, it is found that:
\begin{equation}
\dive \vec{v} = 2\frac{v_r}{r} \label{div}
\end{equation}
The Alfv\'enic perturbations do have a compression. The geometric interpretation is presented in the main body of the paper and in fig. \ref{curve}. The factor $2$ comes from the fact that we require an initial equilibrium state before perturbing with waves. The curvature of field lines exerts a force that must be balanced by a magnetic or thermal pressure gradient. This additional force is equal and opposite to the curvature force in the equilibrium state but acts together with the curvature in the wave equation to strengthen the compression.  

Remembering that the viscous dissipation does not depend exactly on the compression, we need to determine the quantity:
\begin{equation}
D = \dive \vec{v} -3 \partial_\parallel v_\parallel
\end{equation}
From eq. \ref{vtheta}, we get the parallel contribution:
\begin{equation}
\partial_\parallel v_\parallel = \frac{im}{r} v_\theta = \frac{m^2}{r^2}\frac{c_\mathrm{s}^2}{\omega^2} \dive\vec{v}
\end{equation}
so that
\begin{equation}
D = \left(1-3\frac{c_\mathrm{s}^2}{v_\mathrm{A}^2} \right) \dive \vec{v} = 2\left(1-3\frac{c_\mathrm{s}^2}{v_\mathrm{A}^2} \right) \frac{v_r}{r}
\end{equation}
From eq. \ref{vtheta} and \ref{vz}, we see that the general Aflv\'en waves keep their usual polarization: when $k_z=0$, if $v_\theta$ and $v_z$ remain finite, then $v_r$ must vanish, so that we always have: 
\begin{equation}
\vec{k}_\perp.\vec{v}_\perp = 0
\end{equation}
As a consequence, when the Alfv\'en waves are polarized perpendicular to the curvature direction, the curvature of the field lines has no effect on the compression and there is no associated viscous dissipation. However, the Galactic molecular clouds excite waves polarized almost isotropically in every direction, so that the typical perturbed radial velocity can be estimated of the order of the cloud speed $v_\mathrm{c}$.
Finally, defining a local radius of curvature $R_\mathrm{c}$ for the field lines closest to the cloud, we get in order of magnitude:
\begin{equation}
D \approx 2 \left(1-3\frac{c_\mathrm{s}^2}{v_\mathrm{A}^2}\right) \frac{v_\mathrm{c}}{R_\mathrm{c}} 
\end{equation}
At low-$\beta$, $D$ is dominated by the perpendicular divergence of the flow whereas at high-$\beta$, it is dominated by the parallel contribution that may become very large for weak fields. When $c_\mathrm{s}^2/v_\mathrm{A}^2\approx 1/3$, $D$ vanishes and so does the dissipation.

\end{document}